\documentclass{article}[12pt]
\begin{document}

\begin{center}

{\Large \bf THE AMPLITUDE OF LEPTON-PAIR PRODUCTION IN

RELATIVISTIC ION COLLISIONS BEYOND THE\\
\medskip

PERTURBATIVE THEORY}

\end{center}

\begin{center}
{\large \bf A.N. Sissakian, A.V. Tarasov, H.T. Torosyan\footnote{On leave of
absence from Yerevan Physics Institute} and O.O.
Voskresenskaya}\\
\end{center}

\begin{center}
{\small \it Joint Institute for Nuclear Research, 141980 Dubna, Moscow
region, Russia}\\
\end{center}

\noindent
{\bf ABSTRACT}\\
\noindent
{\small We investigate the structure of some first terms of Watson series
representing the amplitude of lepton-pair production in the ion
collisions.
It is shown that infrared instabilities of individual terms trend to
cancel each other providing the infrared stability of whole
amplitude.}

\large

\section{Introduction}

The structure of the amplitude of the reaction
\begin{equation}
Z_1  + Z_2  \to Z_1  + Z_2  + e^ +   + e^ -
\end{equation}
beyond the Born approximation was widely discussed in literature in
past few years [1-12].

It is generally believed that it becomes extremely simple in
ultrarelativistic limit due to practically full contraction of
electromagnetic fields of colliding ions.

The idea to use solution of Dirac equation in the superposition of
these fields for deriving of amplitude of reaction (1) was dominant
in literature for a long time. The extra simple solution of problem
under the consideration obtained in this way have been reported in
several papers [1-3]. Later it have been proved [5,12] that this
solution is incorrect. Detailed discussion of this issue can be
found in [7].

The another approach to this problem
[8,9,11,12] is based on the exploiting of more familiar technique
of Feynman diagrams ($FD$) for the systematic perturbation
calculations of the corrections to the Born approximation result.
The authors [8,9] have proceeded in the summation of the
contributions of the simplest infinite subsets of FD to
$e^{+}e^{-}$-production amplitude, but met serious
difficulties while attempting to  generalize these results to the
more complicated cases.

Partly this problem have been solved in the recent paper [13], where
the Watson type representation of amplitude of reaction (1) in terms
of amplitudes of $e^{\pm}- Z_{1(2)}$-scattering was derived.
The further simplification of this amplitude achieved when one takes
into account the requirement of its infrared stability can be.

\section{Watson series and the problem of infrared stability}

The amplitude of reaction (1) reads
\begin{equation}
A = \overline u \left( {p_2 } \right)T\left( {p_2 , - p_1 }
\right)v\left( {p_1 } \right)
\end{equation}
where $p_1$ and $p_2$ are the four momenta of positron and electron
respectively; $v(p_1)$ and   $u(p_2)$ are their bispinors and the
matrix $T(p_2,-p_1)$ is represented by Watson series
\begin{equation}
\label{trivial}
\begin{array}{l}
  T\left( {p_2 , - p_1 } \right) = T_1 \left( {p_2 , - p_1 } \right)
+ T_2 \left( {p_2 , - p_1 } \right) \\\\
-\int {d^4 kT_1\left({p_2 ,k} \right)} G\left( k \right)T_2
\left({k, - p_1 } \right) - \int {d^4 kT_2 \left( {p_2 ,k}
\right)} G\left( k \right)T_1 \left( {k, - p_1 } \right) \\\\
+ \int{d^4 k_1 d^4 k_2 T_1 \left( {p_2 ,k_1 } \right)} G\left(
{k_1 } \right)T_2 \left( {k_1 ,k_2 } \right)G\left( {k_2 }
\right)T_1 \left( {k_2 , - p_1 } \right) \\\\
+ \int {d^4 k_1 d^4 k_2 T_2 \left( {p_2 ,k_1 } \right)} G\left(
{k_1 } \right)T_1 \left( {k_1 ,k_2 } \right)G\left( {k_2 }
\right)T_2
\left( {k_2 , - p_1 } \right) +  \ldots \\
\end{array}
\end{equation}
or in the short notation
\begin{equation}
\label{trivial}
\begin{array}{l}
  T = T_1  + T_2  - T_1  \otimes G \otimes T_2  - T_2  \otimes G \otimes T_1  
\\
\\
~~~~~~+ T_1  \otimes G \otimes T_2  \otimes G \otimes T_1 + T_2  \otimes G \otimes T_1  \otimes G \otimes T_2  +  \ldots  \\
\end{array}
\end{equation}
above
\begin{equation}
\label{trivial}
\begin{array}{l}
G\left( k \right) = \frac{1}{{\left( {2\pi } \right)^4
}}\frac{{\hat k + m}}{{k^2  - m^2  + i0}},
\\\\
\end{array}
\end{equation}
\begin{equation}
\label{trivial}
\begin{array}{l}
 T_{^1 } \left( {p,p'} \right) = \left( {2\pi } \right)^2 \delta \left( {p_ +   - p'_ +  } \right)\left[ {\theta \left( {p_ +  } \right)f_1^{^{\left(  +  \right)} } \left( {\overrightarrow p _{_T }  - \overrightarrow {p'} _{_T } } \right) - \theta \left( { - p_ +  } \right)f_1 ^{\left(  -  \right)} \left( {\overrightarrow p _{_T }  - \overrightarrow {p'} _{_T } } \right)} \right]\gamma _ +  ,
 \\\\
 T_{^2 } \left( {p,p'} \right) = \left( {2\pi } \right)^2 \delta \left( {p_ -   - p'_ -  } \right)\left[ {\theta \left( {p_ -  } \right)f_2^{^{\left(  +  \right)} } \left( {\overrightarrow p _{_T }  - \overrightarrow {p'} _{_T } } \right) - \theta \left( { - p_ -  } \right)f_2 ^{\left(  -  \right)} \left( {\overrightarrow p _{_T }  - \overrightarrow {p'} _{_T } } \right)} \right]\gamma _ -  ,
 \\\\
 p_ \pm   = p_0  \pm p_z , \\\\
 \gamma _ \pm   = \gamma _0  \pm \gamma _z ,
\\\\
\hat k = k_\mu  \gamma _\mu;
\end{array}
\end{equation}

\begin{equation}
\label{trivial}
\begin{array}{l}
 f_1 ^{\left(  \pm  \right)} \left( {\overrightarrow q } \right) = \frac{i}{{2\pi }}\int {d^2 b} e^{i\overrightarrow q \overrightarrow b } \left[ {1 - S_1 ^{\left(  \pm  \right)} \left( {\overrightarrow b ,\overrightarrow B _1 } \right)} \right],
 \\\\
 f_2 ^{\left(  \pm  \right)} \left( {\overrightarrow q } \right) = \frac{i}{{2\pi }}\int {d^2 b} e^{i\overrightarrow q \overrightarrow b } \left[ {1 - S_2 ^{\left(  \pm  \right)} \left( {\overrightarrow b ,\overrightarrow B _2 } \right)} \right],
 \\\\
 \overrightarrow q  = \overrightarrow p _{_T }  - \overrightarrow {p'} _{_T } , \\
 \end{array}
\end{equation}

\begin{equation}
\label{trivial}
\begin{array}{l}
 S_1 ^{\left(  \pm  \right)} \left( {\overrightarrow b ,\overrightarrow B _1 } \right) = e^{ \pm i\chi _1 \left( {\overrightarrow b ,\overrightarrow B _1 } \right)} ,
 \\\\
 S_2 ^{\left(  \pm  \right)} \left( {\overrightarrow b ,\overrightarrow B _2 } \right) = e^{ \pm i\chi _2 \left( {\overrightarrow b ,\overrightarrow B _2 } \right)} ,
 \\\\
 \end{array}
\end{equation}

\begin{equation}
\label{trivial}
\begin{array}{l}
 \chi _1 \left( {\overrightarrow b ,\overrightarrow B _1 } \right) = e\int\limits_{ - \infty }^\infty  {\Phi _1 \left( {\sqrt {\left( {\overrightarrow b  - \overrightarrow B _1 } \right)^2  + z^2 } } \right)} dz,
 \\\\
 \chi _2 \left( {\overrightarrow b ,\overrightarrow B _2 } \right) = e

 \int\limits_{ - \infty }^\infty  {\Phi _2 \left( {\sqrt {\left( {\overrightarrow b  - \overrightarrow B _2 } \right)^2  + z^2 } } \right)} dz.
 \\\\
 \end{array}
\end{equation}
Here $\Phi _{1,2}$(r) are the Coulomb potentials of ions $Z_1$ and
$Z_2$ respectively, ${\overrightarrow B _1 }$ and ${\overrightarrow
B _2 }$ are their impact parameters.

Since integrals defining the phase shifts $\chi _{1,2}$ are
divergent in the unscreened case, the infinitesimal screening
parameters $\lambda _{1,2}$ are usually introduced as
\begin{equation}
\label{trivial}
\begin{array}{l}
 \Phi _1 \left( r \right) = \mathop {\lim }\limits_{\lambda _1  \to 0} 
\frac{{Z_1 e\exp \left( { - \lambda _1 r} \right)}}{r}
 \\\\
 \Phi _2 \left( r \right) = 
\mathop {\lim }\limits_{\lambda _2  \to 0} 
\frac{{Z_2 e\exp \left( { - \lambda _2 r} \right)}}{r}
 \\\\
 \end{array}
\end{equation}
to make $\chi _{1,2}$ finite.

If some calculated quantity, depending on $\lambda _{1,2}$, approach
the definite limit when $\lambda _{1,2}$ are turned to zero, then
this quantity is called infrared stable. In opposite case it
is known as infrared unstable.

Introduced above  S -matrices  $S_{1(2)}$ are infrared unstable
themselves, but their products
\[
S_1 ^{\left(  +  \right)} \left( {\overrightarrow b
,\overrightarrow B _1 } \right)S_1 ^{\left(  -  \right)} \left(
{\overrightarrow {b'} ,\overrightarrow B _1 } \right)
\]
and
\[
S_2 ^{\left(  +  \right)} \left( {\overrightarrow b
,\overrightarrow B _2 } \right)S_2 ^{\left(  -  \right)} \left(
{\overrightarrow {b'} ,\overrightarrow B _2 } \right)
\]
are infrared stable.

Each term of Watson expansion contains infrared unstable products
of $S_{1,2} ^{\left(  \pm  \right)}$-matrices, 
but only some of these terms contain also the
infrared stable contributions of them. Consequently, each term of
Watson expansion is infrared unstable itself. But sum of all these
terms, representing the amplitude (2), must be infrared stable.

This means that infrared unstable parts of different terms of
Watson expansion must mutually cancel each other.
To see this, we need make some calculations.

\section{The simplest infrared stable
     $e^{+}e^{-}$-production amplitude}

Let us represent the amplitude $A$ in the form
\begin{equation}
\label{trivial}
A = A_1  + A_2  - A_{12}  - A_{21}  + A_{121}  +
A_{212}  - A_{1212}  - A_{2121}  +  \ldots
\end{equation}
where, say,  $A_{121}$ corresponds to the term ${T_1  \otimes G
\otimes T_2  \otimes G \otimes T_1}$   in the Watson expansion
(4).

The simple dependence of ${T_{^{1,2} } \left( {p,p'} \right) }$ on
light-cone components of momenta $p,p^{\prime}$
 allows to make explicitly
all integrations over light-cone components of intermediate momenta
with following results:

\begin{equation}
\label{trivial}
\begin{array}{l}
 {\rm{A}}_{{\rm{12}}}  = \frac{1}{2}\int {d^2 k} a_{12} \left( {\vec k} 
\right)f_1^{\left(  +  \right)} \left( {\vec q_1 } \right)f_2^{\left(  -  
\right)} \left( {\vec q_2 } \right),
 \\\\
 {\rm{A}}_{{\rm{21}}}  = \frac{1}{2}\int {d^2 k} a_{21} \left( {\vec k} 
\right)f_2^{\left(  +  \right)} \left( {\vec q_1 } \right)f_1^{\left(  -  
\right)} \left( {\vec q_2 } \right), \\
 \end{array}
\end{equation}
where
\begin{eqnarray}
\label{trivial}
a_{12} \left( {\vec k} \right)&=& \frac{{\bar u\left( {p_2 } \right)
\gamma _ +  \nu \left( {\vec k} \right)\gamma _ -  
\upsilon \left( {p_1 } \right)}}{{p_{2 + } p_{1 - }  
+ \mu \left( {\vec k} \right)}},\nonumber\\
 a_{21} \left( {\vec k} \right)&=&\frac{{\bar u\left( {p_2 } \right)
\gamma _ -  \nu \left( {\vec k} \right)\gamma _ +  
\upsilon \left( {p_1 } \right)}}{{p_{2 - } p_{1 + }  + 
\mu \left( {\vec k} \right)}},
\end{eqnarray} 
$$\nu \left( {\vec k} \right)= m - \vec \gamma _T \vec k,\quad
 \mu \left( {\vec k} \right) = m^2  + \vec k^2,$$
$$ \vec q_1 = \vec p_{2T}  - \vec k,\qquad
 \vec q_2 = \vec p_{1T}  + \vec k;$$

\newpage
\begin{eqnarray}
A_{121} & = & \frac{1}{(4\pi)^2}
\int d^2 k_1 d^2 k_2 \Big\{ a_{121} ^{( + )}
\left( \vec k_1 ,\vec k_2  \right)f_1^{( + )}
(\vec q_1 )f_2^{(  + )} ( \vec q_2)f_1^{(  - )}(\vec q_3)\nonumber\\ 
&&+a_{121} ^{(-)} \left( \vec k_1 ,\vec k_2 \right)
f_1^{(  +  )} (\vec q_1)f_2^{(  -  )} ( \vec q_2 )
f_1^{(  -  )} \left( \vec q_3 \right)\Big\},\nonumber\\ 
A_{212}&=&\frac{1}{(4\pi)^2}\int d^2 k_1 d^2 k_2  \Big\{
a_{212}^{(  + )} \left(\vec k_1 ,\vec k_2\right)f_2^{(+)}\left(\vec q_1\right)
f_1^{(  +  )} \left(\vec q_2\right)f_2^{(  -  )} \left( \vec q_3 
\right)\nonumber\\
&&+ a_{212}^{(  -  )} \left(\vec k_1 ,\vec k_2\right)
f_2^{\left(  +  \right)} \left( {\vec q_1 }
\right)f_1^{\left(  -  \right)} \left( {\vec q_2 }
\right)f_2^{(  -  )} \left(\vec q_3\right)
\Big\},
\end{eqnarray}
where
\begin{eqnarray}
\label{trivial}
a_{121} ^{\left(  \pm  \right)} \left( {\vec k_1 ,\vec k_2 } \right)& = & 
-a_{121} \left( {\vec k_1 ,\vec k_2 } \right)\left\{ { \ln \left[ {\frac{{\mu \left( {\vec k_1 } \right)p_{1 + } }}
{{\mu \left( {\vec k_2 } \right)p_{2 + } }}} \right] \pm \pi i}
\right\},\nonumber\\
a_{121} \left( {\vec k_1 ,\vec k_2 } \right)& = &
\frac{{\bar u\left( {p_2 } \right)\gamma _ +  \nu \left( {\vec k_1 } \right)\gamma _ -  \nu \left( {\vec k_2 } \right)\gamma _ +  \upsilon \left( {p_1 } \right)}}
{{p_{1 + } \mu \left( {\vec k_1 } \right) + p_{2 + } \mu \left(
{\vec k_2 } \right)}},\nonumber\\
a_{212} ^{\left(  \pm  \right)} \left( {\vec k_1 ,\vec k_2 } \right)& =&
 -a_{212} \left( {\vec k_1 ,\vec k_2 } \right)\left\{ { \ln \left[ {\frac{{\mu \left( {\vec k_1 } \right)p_{1 - } }}
{{\mu \left( {\vec k_2 } \right)p_{2 - } }}} \right] \pm \pi i}
\right\},\nonumber\\
a_{212} \left( {\vec k_1 ,\vec k_2 } \right)& = & 
\frac{\bar u(p_2)\gamma_- \nu\left(\vec k_1\right)
\gamma _ + \nu \left(\vec k_2\right)\gamma _- \upsilon \left(p_1\right)}
{p_{1 - }\mu\left(\vec k_1\right) + p_{2-}\mu\left(\vec k_2\right)},
\end{eqnarray}

$$\vec q_1  = 
\vec p_{2T}  - \vec k_1 ,\quad \vec q_2  = \vec k_1  - \vec k_2 ,
\quad\vec q_3  = 
\vec p_{1T}  + \vec k_2 ;$$ 

\begin{eqnarray}
\label{trivial}
A_{1212}  = A_{1212} ^{\left( s
\right)}  + A_{1212} ^{\left( u \right)},
\end{eqnarray}
where
\begin{eqnarray}
\label{trivial}
A_{1212} ^{\left( s \right)} & =& 
\int {d^2 k_1 d^2 k_2 d^2 k_3 } a_{1212} ^{\left( s \right)} 
\left( {\vec k_1 ,\vec k_2 ,\vec k_3 } \right)f_1^{\left(  +  
\right)} \left( {\vec q_1 } \right)f_2^{\left(  +  \right)} 
\left( {\vec q_2 } \right)f_1^{\left(  -  \right)} \left( {\vec q_3 } 
\right)f_2^{\left(  -  \right)} \left( {\vec q_4 } \right), \nonumber\\
A_{1212} ^{\left( u \right)} & = &
\int {d^2 k_1 d^2 k_2 d^2 k_3 } a_{1212} ^{\left( u \right)} 
\left( {\vec k_1 ,\vec k_2 ,\vec k_3 } \right)f_1^{\left(  +  \right)} 
\left( {\vec q_1 } \right)\left[ {f_2^{\left(  +  \right)} 
\left( {\vec q_2 } \right) + f_2^{\left(  -  \right)} \left( {\vec q_2 } 
\right)} \right]\\
&&\times \left[ {f_1^{\left(  +  \right)} \left( {\vec q_3 } \right) 
+ f_1^{\left(  -  \right)} \left( {\vec q_3 } \right)} \right]f_2^{\left(  
-  \right)} \left( {\vec q_4 } \right);\nonumber
\end{eqnarray}
\begin{eqnarray}
\label{trivial}
A_{2121} & = &A_{2121}^{\left( s
\right)}  + A_{2121}^{\left( u \right)},
\end{eqnarray}
where
\begin{eqnarray}
\label{trivial}
A_{2121}^{\left( s \right)} & = &\int {d^2 k_1 d^2 k_2 d^2 k_3 } 
a_{2121}^{\left( s \right)} \left( {\vec k_1 ,\vec k_2 ,\vec k_3 } 
\right)f_2^{\left(  +  \right)} \left( {\vec q_1 } \right)f_1^{\left(  +  
\right)} \left( {\vec q_2 } \right)f_2^{\left(  -  
\right)} \left( {\vec q_3 } \right)f_1^{\left(  -  \right)} 
\left( {\vec q_4 } \right),  \nonumber\\
A_{2121}^{\left( u \right)}&  = &\int {d^2 k_1 d^2 k_2 d^2 k_3 } 
a_{2121}^{\left( u \right)} \left( {\vec k_1 ,\vec k_2 ,\vec k_3 } 
\right)f_2^{\left(  +  \right)} \left( {\vec q_1 } 
\right)\left[ {f_1^{\left(  +  \right)} \left( {\vec q_2 } \right) 
+ f_1^{\left(  -  \right)} \left( {\vec q_2 } \right)} \right]\\
&&  \times \left[ {f_2^{\left(  +  \right)} \left( {\vec q_3 } \right) 
+ f_2^{\left(  -  \right)} \left( {\vec q_3 } \right)} \right]f_1^{\left(  
-  \right)} \left( {\vec q_4 } \right), \nonumber
\end{eqnarray}

\begin{eqnarray}
\label{trivial}
a_{1212} ^{\left( s \right)} \left( {\vec k_1 ,\vec k_2 ,\vec k_3 } \right)
&=& a_{1212} \left( {\vec k_1 ,\vec k_2 ,\vec k_3 } \right)
\left\{ { - \ln \left[ {\frac{{\mu \left( {\vec k_1 } \right)
\mu \left( {\vec k_3 } \right)}}
{{\mu \left( {\vec k_2 } \right)p_{2 + } p_{1 - } }}} \right] - i
\pi } \right\},\nonumber\\
a_{1212} ^{\left( u \right)} \left( {\vec k_1 ,\vec k_2 ,\vec k_3 } \right) 
&=& a_{1212} \left( {\vec k_1 ,\vec k_2 ,\vec k_3 } \right)\frac{1}
{3}\left\{ { \ln ^2 \left[ {\frac{{\mu \left( {\vec k_1 } \right)\mu
\left( {\vec k_3 } \right)}} {{\mu \left( {\vec k_2 } \right)p_{2 +
} p_{1 - } }}} \right] + \pi ^2 } \right\},\nonumber\\
a_{2121} ^{\left( s \right)} \left( {\vec k_1 ,\vec k_2 ,\vec k_3 } \right) 
&=& a_{2121} \left( {\vec k_1 ,\vec k_2 ,\vec k_3 } \right)\left\{ { - \ln \left[ {\frac{{\mu \left( {\vec k_1 } \right)\mu \left( {\vec k_3 } \right)}}
{{\mu \left( {\vec k_2 } \right)p_{2 - } p_{1 + } }}} \right] - i
\pi } \right\},\nonumber\\
a_{2121} ^{\left( u \right)} \left( {\vec k_1 ,\vec k_2 ,\vec k_3 } \right) 
&=& a_{2121} \left( {\vec k_1 ,\vec k_2 ,\vec k_3 } \right)\frac{1}
{3}\left\{ {\ln ^2 \left[ {\frac{{\mu \left( {\vec k_1 } \right)\mu
\left( {\vec k_3 } \right)}} {{\mu \left( {\vec k_2 } \right)p_{2 -
} p_{1 + } }}} \right] + \pi ^2 } \right\},\nonumber \\
a_{1212} \left( {\vec k_1 ,\vec k_2 ,\vec k_3 } \right)& =& \frac{{\bar u\left( {p_2 } \right)\gamma _ +  \nu \left( {\vec k_1 } \right)\gamma _ -  \nu \left( {\vec k_2 } \right)\gamma _ +  \nu \left( {\vec k_3 } \right)\gamma _ -  \upsilon \left( {p_1 } \right)}}
{{\mu \left( {\vec k_1 } \right)\mu \left( {\vec k_3 } \right) + \mu
\left( {\vec k_2 } \right)p_{2 + } p_{1 - } }},\nonumber\\
a_{2121} \left( {\vec k_1 ,\vec k_2 ,\vec k_3 } \right)& =& 
\frac{{\bar u\left( {p_2 } \right)\gamma _ -  \nu \left( {\vec k_1 } \right)\gamma _ +  \nu \left( {\vec k_2 } \right)\gamma _ -  \nu \left( {\vec k_3 } \right)\gamma _ +  \upsilon \left( {p_1 } \right)}}
{{\mu \left( {\vec k_1 } \right)\mu \left( {\vec k_3 } \right) + \mu
\left( {\vec k_2 } \right)p_{2 - } p_{1 + } }}, 
\end{eqnarray}

$$
\vec q_1  = \vec p_{2T}  - \vec k_1 ,\quad \vec q_2  = \vec k_1  - \vec k_2 ,
\quad\vec q_3  = \vec k_2  - \vec k_3 ,~~\vec q_4  = \vec p_{1T}  + \vec k_3. 
$$

Substituting in these expressions $ f_{1,2}^{\left( \pm
\right)}$-amplitudes in the form
\begin{eqnarray}
\label{trivial}
 f_1^{\left(  \pm  \right)} \left( {\vec q} \right)& =& 
2\pi i\left( {\delta \left( {\vec q} \right) - 
S_1^{\left(  \pm  \right)} \left( {\vec q,\vec B_1 } \right)} \right),\\
f_2^{\left(  \pm  \right)} \left( {\vec q} \right)& =& 
2\pi i\left( {\delta \left( {\vec q} \right) - 
S_2^{\left(  \pm  \right)} \left( {\vec q,\vec B_2 } \right)} \right), \nonumber
\end{eqnarray}
where
\begin{eqnarray}
\label{trivial}
S_1^{\left(  \pm  \right)} \left( {\vec q,\vec B_1 } \right)& =& \frac{1}
{{\left( {2\pi } \right)^2 }}\int {d^2 b\exp (i\vec q\vec b)~}
S_1^{\left(  \pm  \right)} \left( {\vec b,\vec B_1 } \right),  \\
S_2^{\left(  \pm  \right)} \left( {\vec q,\vec B_2 } \right)& =& \frac{1}
{{\left( {2\pi } \right)^2 }}\int {d^2 b\exp (i\vec q\vec b)~}
S_2^{\left(  \pm  \right)} \left( {\vec b,\vec B_2 } \right), \nonumber
\end{eqnarray}
one obtains the amplitude (2) in terms of products 
$S_{1(2)}^{\left( \pm \right)}$-matrices.

Using the following identities
\begin{eqnarray}
\label{trivial}
\gamma _ \pm  \nu \left( { - \vec p_{1T} } \right)\gamma _ \mp  \upsilon 
\left( {p_1 } \right)& \equiv&  - 2p_{1 \pm } \gamma _ \pm  \upsilon 
\left( {p_1 } \right),  \nonumber\\
\bar u\left( {p_2 } \right)\gamma _ \pm  \nu \left( {\vec p_{2T} } \right)
\gamma _ \mp  & \equiv& 2p_{2 \pm } \bar u\left( {p_2 } \right)\gamma _ \pm ,\\
\gamma _ \pm  \nu \left( {\vec k} \right)\gamma _ \mp  \nu 
\left( {\vec k} \right)\gamma _ \pm  & \equiv & 4\mu \left( {\vec k} \right)
\gamma _ \pm, \nonumber
\end{eqnarray}
one can easily check that the sum
\begin{equation}
\label{trivial}
\tilde A  = A_{12}  + A_{21}  -  A_{121}  - A_{212}  + A_{1212}^{( s )}  +
A_{2121}^{( s ) }
\end{equation}
is infrared stable and reads
\begin{eqnarray}
\label{trivial}
\tilde A&=&\frac{{i \pi}} {4}\int {d^2 k_1 d^2 k_2 } d^2
k_3 a_{1212} ^{\left( s \right)} \left( {\vec k_1 ,\vec k_2 ,\vec
k_3 }\right)\\
&&\times\left\{ {\delta \left( {\vec q_1 } \right)\delta
\left( {\vec q_2 } \right)\delta \left( {\vec q_3 } \right)\delta
\left( {\vec q_4 } \right) - S_1 ^{\left(  +  \right)} \left( {\vec
q_1 ,\overrightarrow B _1 } \right)S_2 ^{\left(  +  \right)} \left(
{\vec q_2 ,\overrightarrow B _2 } \right)S_1 ^{\left(  -  \right)}
\left( {\vec q_3 ,\overrightarrow B _1 } \right)S_2 ^{\left(  -
\right)} \left( {\vec q_4 ,\overrightarrow B _1 } \right)} \right\}
\nonumber\\
&&+\frac{{i \pi}} {4}\int {d^2 k_1 d^2 k_2 } d^2 k_3 a_{2121}
^{\left( s \right)} \left( {\vec k_1 ,\vec k_2 ,\vec k_3 }
\right)\nonumber\\
&&\times\left\{ {\delta \left( {\vec q_1 } \right)\delta \left( {\vec
q_2 } \right)\delta \left( {\vec q_3 } \right)\delta \left( {\vec
q_4 } \right) - S_2 ^{\left(  +  \right)} \left( {\vec q_1
,\overrightarrow B _2 } \right)S_1 ^{\left(  +  \right)} \left(
{\vec q_2 ,\overrightarrow B _1 } \right)S_2 ^{\left(  -  \right)}
\left( {\vec q_3 ,\overrightarrow B _2 } \right)S_1 ^{\left(  -
\right)} \left( {\vec q_4 ,\overrightarrow B _1 } \right)} 
\right\}.\nonumber
\end{eqnarray}

Thus, all infrared unstable terms in (24)
cancel each other.

The rest
$$A_{1212}^{\left( u \right)}  + A_{2121}^{\left( u \right)}$$
is of order $(Z_{1}\alpha \cdot Z_{2}\alpha)^3$, as a  $A_{121212}$
and $A_{212121}$, which is beyond of this consideration, and we hope
that its infrared unstable components must be canceled with similar
unstable expressions of higher order terms, like was shown above for
lower order case. We plan to consider this issue elsewhere.

\section{Conclusion}

The Watson-type representation for the amplitude of production of
$e^{+}e^{-}$-pair in the ion-ion collisions can be considered
as a resummation of FD of perturbation theory in more economic form.
The application of the requirement of infrared stability of this
amplitude provides new tool for future simplification.

\section*{Acknowledgments}

The authors gratefully acknowledge fruitful discussions with
S.R. Gevorkyan and E.A. Kuraev.

\end{document}